\documentstyle[prl,eqsecnum,aps]{revtex}

\begin{document}
\author{Jian-Qi Shen \footnote{E-mail address: jqshen@coer.zju.edu.cn}$^{1,2}$, Pan Chen$^{1,2}$ and Hong Mao$^{2}$}
\address{1. Center for Optical
and Electromagnetic Research, State Key Laboratory of \\Modern
Optical Instrumentation,
College of Information Science and Engineering%
\\
2.Zhejiang Institute of Modern Physics and Department of Physics,\\
Zhejiang University, Hangzhou 310027, People$^{,}$s Republic of China}
\date{\today}
\title{Exact time-dependent decoherence factor \\and its adiabatic classical limit}
\maketitle

\begin{abstract}
The present paper finds the complete set of exact solutions of the
general time-dependent dynamical models for quantum decoherence,
by making use of the Lewis-Riesenfeld invariant theory and the
invariant-related unitary transformation formulation. Based on
this, the general explicit expression for the decoherence factor
is then obtained and the adiabatic classical limit of an
illustrative example is discussed. The result ( i.e., the
adiabatic classical limit) obtained in this paper is consistent
with what obtained by other authors, and futhermore we obtain the
more general results concerning the time-dependent non-adiabatic
quantum decoherence. It is shown that the invariant theory is
appropriate for treating both the time-dependent quantum
decoherence and the geometric phase factor.

PACS:  03.65.Ge, 03.65.Bz

Keywords: decoherence factor; invariant theory; time-dependent
decoherence
\end{abstract}
\pacs{PACS: 03.65.Ge 03.65.Bz }

\section{INTRODUCTION}

Solvable models in quantum mechanics enable one to investigate
quantum measurement problems very conveniently\cite{Nakazato}. A
good number of investigators have studied these useful models such
as Hepp-Coleman model \cite{Namiki} and Cini model\cite{Cini}. The
exact solvability of these models often provides physicists with a
clear understanding of the physical phenomena involved and yield
rich physical insights\cite {Namiki2,Nakazato2}. In these works,
the first important step is to obtain the exact solutions of the
Schr\"{o}dinger equation and the time-evolution operator, which
can be applied to the calculation of the decoherence factor and
study of the wavefunction collapse, etc.. Although the exact
solutions and the decoherence of these models have been
extensively studied by many authors, the coefficients in the
Hamiltonians of all these models are merely time-independent (or
partially time-dependent), to the best of our knowledge. In the
present paper, we obtain the explicit time-evolution operator and
the decoherence factor of the general dynamical models where the
Hamiltonians are totally time-dependent.

Time-dependent system is governed by the time-dependent
Schr\"{o}dinger equation. The invariant theory\cite{Lewis}
suggested by Lewis and Riesenfeld in 1969 can solve the
time-dependent Schr\"{o}dinger equation. In 1991, Gao {\it et al}
proposed a generalized invariant theory\cite{Gao1} by introducing
basic invariants, which enable one to find the complete set of
commuting invariants for some time-dependent multi-dimensional
systems\cite {Gao4,Shen,Kim}. We will analyze the general
dynamical model in what follows and then calculate the
time-dependent decoherence factor by making use of these invariant
theories.

\section{EXACT DECOHERENCE FACTOR IN TIME-DEPENDENT DECOHERENCE}

The original Cini model for the correlation between the states of the
measured system and the measuring instrument-detector is built for a
two-level system interacting with the detector. Liu and Sun generalized this
Cini model to an $M$-level system\cite{Liu}. In this paper we further
consturct an general dynamical model for quantum decoherence between the
measuring instrument-detector and the measured system where $A_{+},A_{-}$
and $A$ denote the measuring instrument-detector; and $\omega _{i}(t)$
represents the energy parameter of a certain state $\left| i\right\rangle $
of the multi-level measured system and $\theta _{i}(t)$ and $\phi _{i}(t)$
are coupling coefficients of interaction of the measuring
instrument-detector with the measured system. The Hamiltonian which
describes the interaction between the state $\left| i\right\rangle $ of the
multi-level measured system and the measuring instrument-detector is then
given as follows:

\begin{eqnarray}
H_{i}(t) &=&\omega _{i}(t)\{\frac{1}{2}\sin \theta _{i}(t)\exp [-i\phi
_{i}(t)]A_{+}  \nonumber \\
&&+\frac{1}{2}\sin \theta _{i}(t)\exp [i\phi _{i}(t)]A_{-}+\cos \theta
_{i}(t)A\}  \label{eq18}
\end{eqnarray}
with $A,A_{+-}$ and $A$ satisfying the general commuting relations of a Lie
algebra

\begin{equation}
\lbrack A_{+},A_{-}]=nA,\quad \lbrack A,A_{+}]=mA_{+},\quad \lbrack
A,A_{-}]=-mA_{-},  \label{eq19}
\end{equation}
where $m,-m$ and $n$ are structure constants of this Lie algebra.
Time evolution of this dynamical model is governed by the
Schr\"{o}dinger equation (in the unit $\hbar=1$)

\begin{equation}
i\frac{\partial \left| \Psi _{i}(t)\right\rangle _{s}}{\partial t}%
=H_{i}(t)\left| \Psi _{i}(t)\right\rangle _{s}.  \label{eq17}
\end{equation}

According to the Lewis-Riesenfeld invariant theory, an operator $I(t)$ that
agrees with the following invariant equation\cite{Lewis}

\begin{equation}
\frac{\partial I_{i}(t)}{\partial t}+\frac{1}{i}[I_{i}(t),H_{i}(t)]=0
\label{eq20}
\end{equation}
is called an invariant whose eigenvalue is time-independent, i.e.,

\begin{equation}
I_{i}(t)\left| \lambda ,i,t\right\rangle =\lambda \left| \lambda
,i,t\right\rangle _{i},\quad \frac{\partial \lambda }{\partial t}=0.
\label{eq21}
\end{equation}
It is seen from Eq. (\ref{eq20}) that $I_{i}(t)$ is the linear combination
of $A_{+},A_{-}$and $A$\bigskip\ and may be generally written\qquad

\begin{equation}
I_{i}(t)=y\{\frac{1}{2}\sin a_{i}(t)\exp [-ib_{i}(t)]A_{+}+\frac{1}{2}\sin
a_{i}(t)\exp [ib_{i}(t)]A_{-}\}+\cos a_{i}(t)A,  \label{eq22}
\end{equation}
where the constant $y$ will be determined below. Substitution of (\ref{eq22}%
) into Eq. (\ref{eq20}) yields

\begin{eqnarray}
y\exp (-ib_{i})(\dot{a}_{i}\cos a_{i}-i\dot{b}_{i}\sin a_{i})-im\omega
_{i}[\exp (-i\phi _{i})\cos a_{i}\sin \theta _{i}-y\exp (-ib_{i})\sin
a_{i}\cos \theta _{i}] &=&0, \\
\dot{a}_{i}+\frac{ny}{2}\omega _{i}\sin \theta _{i}\sin (b_{i}-\phi _{i})
&=&0.  \label{eq23}
\end{eqnarray}
where dot denotes the time derivative. The time-dependent parameters $a_{i}$
and $b_{i}$ are determined by these two auxiliary equations.

It is easy to verify that the particular solution $\left| \Psi
_{i}(t)\right\rangle _{s}$ of the Schr\"{o}dinger equation can be expressed
in terms of the eigenstate $\left| \lambda ,i,t\right\rangle $ of the
invariant $I_{i}(t),$ namely,

\begin{equation}
\left| \Psi _{i}(t)\right\rangle _{s}=\exp [\frac{1}{i}\varphi
_{i}(t)]\left| \lambda ,i,t\right\rangle  \label{eq24}
\end{equation}
with
\begin{equation}
\varphi _{i}(t)=\int_{0}^{t}\left\langle \lambda ,i,t^{^{\prime
}}\right| [H_{i}(t^{^{\prime }})-i\frac{\partial }{\partial
t^{^{\prime }}}]\left| \lambda ,i,t^{^{\prime }}\right\rangle {\rm
d}t^{^{\prime }}.
\end{equation}
The physical meanings of $\int_{0}^{t}\left\langle \lambda
,i,t^{^{\prime }}\right| H_{i}(t^{^{\prime }})\left| \lambda
,i,t^{^{\prime }}\right\rangle {\rm d}t^{^{\prime }}$ and
$\int_{0}^{t}\left\langle \lambda ,i,t^{^{\prime }}\right|
-i\frac{\partial }{\partial t^{^{\prime }}}\left| \lambda
,i,t^{^{\prime }}\right\rangle {\rm d}t^{^{\prime }}$ are
dynamical and geometric phase, respectively.

Since the expression (\ref{eq24}) is merely a formal solution of the
Schr\"{o}dinger equation, in order to get the explicit solutions we make use
of the invariant-related unitary transformation formulation\cite{Gao1} which
enables one to obtain the complete set of exact solutions of the
time-dependent Schr\"{o}dinger equation (\ref{eq17}). In accordance with the
invariant-related unitary transformation method, the time-dependent unitary
transformation operator is often of the form

\begin{equation}
V_{i}(t)=\exp [\beta _{i}(t)A_{+}-\beta _{i}^{\ast }(t)A_{-}]  \label{eq25}
\end{equation}
with $\beta _{i}(t)=-\frac{a_{i}(t)}{2}x\exp [-ib_{i}(t)],\quad
\beta _{i}^{\ast }(t)=-\frac{a_{i}(t)}{2}x\exp [ib_{i}(t)].$ By
making use of the Glauber formula, lengthy calculation yields

\begin{eqnarray}
I_{iV} &=&V_{i}^{\dagger }(t)I_{i}(t)V_{i}(t)=\{\frac{y}{2}\exp
(-ib_{i})\sin a_{i}\cos [(\frac{mn}{2})^{\frac{1}{2}}a_{i}x]  \nonumber \\
&&-\frac{(\frac{mn}{2})^{\frac{1}{2}}}{n}\exp (-ib_{i})\cos a_{i}\sin [(%
\frac{mn}{2})^{\frac{1}{2}}a_{i}x]\}A_{+}  \nonumber \\
&&+\{\frac{y}{2}\exp (ib_{i})\sin a_{i}\cos [(\frac{mn}{2})^{\frac{1}{2}%
}a_{i}x]  \nonumber \\
&&-\frac{(\frac{mn}{2})^{\frac{1}{2}}}{n}\exp (ib_{i})\cos a_{i}\sin [(\frac{%
mn}{2})^{\frac{1}{2}}a_{i}x]\}A_{-}  \nonumber \\
&&+\{\cos a_{i}\cos [(\frac{mn}{2})^{\frac{1}{2}}a_{i}x]+\frac{(\frac{mn}{2}%
)^{\frac{1}{2}}}{m}y\sin a_{i}\sin [(\frac{mn}{2})^{\frac{1}{2}}a_{i}x]\}A.
\end{eqnarray}
It can be easily seen that when
\begin{equation}
y=\frac{m}{(\frac{mn}{2})^{\frac{1}{2}}},\quad x=\frac{1}{(\frac{mn}{2})^{%
\frac{1}{2}}},
\end{equation}
one may derive that $I_{iV}=A$ which is time-independent. Thus the
eigenvalue equation of the time-independent invariant $I_{iV}$ may
be written in the form

\begin{equation}
I_{iV}\left| \lambda \right\rangle =\lambda \left| \lambda \right\rangle
,\quad \left| \lambda \right\rangle =V_{i}^{\dagger }(t)\left| \lambda
,i,t\right\rangle .  \label{eq27}
\end{equation}
In the meanwhile, by the aid of Baker-Campbell-Hausdorff
formula\cite{Wei}, one can arrive at
\begin{eqnarray}
V_{i}^{\dagger }(t)H_{i}(t)V_{i}(t) &=&\{\omega
_{i}[\frac{1}{2}\sin \theta _{i}\exp (-i\phi
_{i})-\frac{\sqrt{\frac{mn}{2}}}{n}\exp (-ib_{i})\cos
\theta _{i}\sin (\sqrt{\frac{mn}{2}}a_{i}x)]   \nonumber \\
&&+\frac{1}{2}\omega _{i}\exp (-ib_{i})\sin \theta _{i}\cos
(b_{i}-\phi _{i})[\cos (\sqrt{\frac{mn}{2}}a_{i}x)-1]\}A_{+} \nonumber \\
&&+\{\omega _{i}[\frac{1}{2}\sin \theta _{i}\exp (i\phi _{i})-\frac{\sqrt{%
\frac{mn}{2}}}{n}\exp (ib_{i})\cos \theta _{i}\sin (\sqrt{\frac{mn}{2}}%
a_{i}x)] \nonumber \\
&&+\frac{1}{2}\omega _{i}\exp (ib_{i})\sin
\theta _{i}\cos
(b_{i}-\phi _{i})[\cos (\sqrt{\frac{mn}{2}}a_{i}x)-1]\}A_{-}  \nonumber \\
&&+\omega _{i}[\cos \theta _{i}\cos (\sqrt{\frac{mn}{2}}a_{i}x)+\frac{\sqrt{%
\frac{mn}{2}}}{m}\sin \theta _{i}\cos (b_{i}-\phi _{i})\sin (\sqrt{\frac{mn}{%
2}}a_{i}x)]A,    \label{eq027}
\end{eqnarray}
which is related to the dynamical phase; and
\begin{eqnarray}
V_{i}^{\dagger }(t)i\frac{\partial V_{i}(t)}{\partial t} &=&[-\frac{i}{2}%
\dot{a}_{i}x\exp (-ib_{i})-\frac{1}{2}\frac{1}{\sqrt{\frac{mn}{2}}}\dot{b}%
_{i}\exp (-ib_{i})\sin (\sqrt{\frac{mn}{2}}a_{i}x)]A_{+} \nonumber \\
&&+[\frac{i}{2}\dot{a}_{i}x\exp (ib_{i})-\frac{1}{2}\frac{1}{\sqrt{\frac{mn}{%
2}}}\dot{b}_{i}\exp (ib_{i})\sin (\sqrt{\frac{mn}{2}}a_{i}x)]A_{-} \nonumber \\
&&-\frac{\dot{b}_{i}}{m}[1-\cos (\sqrt{\frac{mn}{2}}a_{i}x)]A,
\label{eq028}
\end{eqnarray}
which is related to the geometric phase. It follows from Eq.
(\ref{eq027}) and Eq. (\ref{eq028}) that, under the transformation
$V(t),$ the Hamiltonian $H(t)$ can be changed into

\begin{eqnarray}
H_{iV}(t) &=&V_{i}^{\dagger }(t)H_{i}(t)V_{i}(t)-V_{i}^{\dagger }(t)i\frac{%
\partial V_{i}(t)}{\partial t}  \nonumber \\
&=&\{\omega _{i}[\cos a_{i}\cos \theta _{i}+\frac{(\frac{mn}{2})^{\frac{1}{2}%
}}{m}\sin a_{i}\sin \theta _{i}\cos (b_{i}-\phi _{i})]  \nonumber \\
&&+\frac{\dot{b}_{i}}{m}(1-\cos a_{i})\}A  \label{eq28}
\end{eqnarray}
Hence, with the help of Eq. (\ref{eq24}) and Eq. (\ref{eq27}), the
particular solution of the Schr\"{o}dinger equation is obtained

\begin{equation}
\left| \Psi _{i}(t)\right\rangle _{s}=\exp [\frac{1}{i}\varphi
_{i}(t)]V_{i}(t)\left| \lambda ,i\right\rangle  \label{eq29}
\end{equation}
with the phase

\begin{eqnarray}
\varphi _{i}(t) &=&\int_{0}^{t}\left\langle \lambda \right|
[V_{i}^{\dagger }(t^{^{\prime }})H_{i}(t^{^{\prime
}})V_{i}(t^{^{\prime }})-V_{i}^{\dagger }(t^{^{\prime
}})i\frac{\partial }{\partial t^{^{\prime }}}V_{i}(t^{^{\prime
}})]\left| \lambda \right\rangle {\rm
d}t^{^{\prime }}  \nonumber \\
&=&\varphi _{id}(t)+\varphi _{ig}(t)  \nonumber \\
&=&\lambda \int_{0}^{t}\{\omega _{i}[\cos a_{i}\cos \theta _{i}+\frac{(\frac{%
mn}{2})^{\frac{1}{2}}}{m}\sin a_{i}\sin \theta _{i}\cos (b_{i}-\phi _{i})]
\nonumber \\
&&+\frac{\dot{b}_{i}}{m}(1-\cos a_{i})\}{\rm d}t^{^{\prime }},
\label{eq30}
\end{eqnarray}
where the dynamical phase is $\varphi _{id}(t)=\lambda
\int_{0}^{t}\omega _{i}[\cos a_{i}\cos \theta
_{i}+\frac{(\frac{mn}{2})^{\frac{1}{2}}}{m}\sin a_{i}\sin \theta
_{i}\cos (b_{i}-\phi _{i})]{\rm d}t^{^{\prime }}$ and the
geometric phase is $\varphi _{ig}(t)=\lambda \int_{0}^{t}\frac{\dot{b}_{i}}{m%
}(1-\cos a_{i}){\rm d}t^{^{\prime }}.$ It is seen that the former
phase is related to the dynamical parameters of the Hamiltonian
such as $\omega _{i},\cos \theta _{i},\sin \theta _{i},$ etc.,
whereas the latter is not immediately related to these parameters.
If the parameter $a_{i}$ is taken to be time-independent, then we
arrive at

\begin{equation}
\varphi _{ig}(T)=\lambda \int_{0}^{T}\frac{\dot{b}_{i}}{m}(1-\cos
a_{i}){\rm d}t^{^{\prime }}=\frac{\lambda }{m}[2\pi (1-\cos
a_{i})],
\end{equation}
where $2\pi (1-\cos a_{i})$ is an expression for the solid angle
over the parameter space of the invariant. This fact shows the
global or topological meanings of the geometric phase $\varphi
_{ig}(t)$\cite{Berry}. The expression (\ref{eq29}) is a particular
exact solution corresponding to $\lambda $ and the general
solutions of the time-dependent Schr\"{o}dinger equation are
easily obtained by using the linear combinations of all these
particular solutions.

Since we have exact solutions of the general time-dependent model,
we can obtain the exact expression for the time-dependent
decoherence factor that is given
\begin{equation}
F_{i,j}(t)=\left\langle \lambda \right| V_{i}^{\dagger
}(t)V_{j}(t)\left| \lambda \right\rangle .  \label{eq200}
\end{equation}
Further calculation yields

\begin{equation}
F_{i,j}(t)=\exp [\frac{n\lambda }{2}(\beta _{i}\beta _{j}^{\ast
}-\beta _{i}^{\ast }\beta _{j})]\left\langle \lambda \right| \exp
[(\beta _{j}-\beta _{i})A_{+}-(\beta _{j}^{\ast }-\beta _{i}^{\ast
})A_{-}]\left| \lambda \right\rangle,
\end{equation}
which is the general expression for the decoherence factor of the
time-dependent dynamical model (\ref{eq18}). Although the expression (\ref
{eq21}) is somewhat complicated, it is just the explicit expression that
does not contain the chronological product.

\section{AN ILLUSTRATIVE EXAMPLE}

To show that (\ref{eq21}) descends to the result familiar to us in
the time-independent (or partially time-dependent) dynamical model
of decoherence, we take into consideration the adiabatic classical
limit of a special dynamical model. Liu and Sun generalized the
original Cini model to an $M$-level system\cite{Liu}. The
Hamiltonian of this generalized model is written

\begin{equation}
H=H_{S}+H_{D}+H_{I},  \label{eq1}
\end{equation}
where $H_{S}$ is the model Hamiltonian of the measured system S with $M$
levels and $H_{D}$ is the free Hamiltonian of the two-boson-state detector
D. They are generally of the forms

\begin{equation}
H_{S}=\sum_{k=1}^{M}E_{k}\left| \Phi _{k}\right\rangle \left\langle \Phi
_{k}\right| ,\quad H_{D}=\omega _{1}a_{1}^{\dagger }a_{1}+\omega
_{2}a_{2}^{\dagger }a_{2}  \label{eq2}
\end{equation}
with the creation and annihilation operators $a_{i}^{\dagger }$ , $a_{i}$
satisfying the following commuting relations

\begin{equation}
\left[ a_{i},a_{j}^{\dagger }\right] =\delta _{ij},\quad \left[ a_{i},a_{j}%
\right] =\left[ a_{i}^{\dagger },a_{j}^{\dagger }\right] =0.  \label{eq3}
\end{equation}
The interaction Hamiltonian $H_{I}$ is given by

\begin{eqnarray}
H_{I} &=&\sum_{n}\left| \Phi _{n}\right\rangle \left\langle \Phi _{n}\right|
(g_{n}a_{1}^{\dagger }a_{2}+g_{n}^{\ast }a_{2}^{\dagger }a_{1})  \nonumber \\
&=&\sum_{n}\left| \Phi _{n}\right\rangle \left\langle \Phi _{n}\right|
(g_{n}J_{+}+g_{n}^{\ast }J_{-})
\end{eqnarray}
with $J_{+}=a_{1}^{\dagger }a_{2},J_{-}=a_{2}^{\dagger }a_{1},J_{3}=\frac{1}{%
2}(a_{1}^{\dagger }a_{1}-a_{2}^{\dagger }a_{2})$ satisfying the commuting
relations $[J_{+},J_{-}]=2J_{3},[J_{3},J_{\pm }]=\pm J_{\pm }.$

It can be seen from the form of the Hamiltonian that both $\left| \Phi
_{k}\right\rangle \left\langle \Phi _{k}\right| $ and $N=\frac{%
a_{1}^{\dagger }a_{1}+a_{2}^{\dagger }a_{2}}{2}$ commute with $H,$namely, $%
\left[ \left| \Phi _{k}\right\rangle \left\langle \Phi _{k}\right|
,H\right] =\left[ N,H\right] =0.$ Hence, a generalized
quasialgebra which enables one to obtain the complete set of exact
solutions of the Schr\"{o}dinger equation can be found by working
in a sub-Hilbert-space corresponding to the particular eigenvalues
of both $\left| \Phi _{k}\right\rangle \left\langle \Phi
_{k}\right| $ and $N$, and then the Hamiltonian can be rewritten
in this sub-Hilbert-space

\begin{equation}
H_{n,k}(t)=E_{k}+g_{k}J_{+}+g_{k}^{\ast }J_{-}+(\omega _{1}-\omega
_{2})J_{3}+n(\omega _{1}+\omega _{2})  \label{eq6}
\end{equation}
with $n$ being the eigenvalue of $N$ and satisfying

\begin{equation}
N\left| n_{1},n_{2}\right\rangle =n\left| n_{1},n_{2}\right\rangle ,\quad n=%
\frac{1}{2}(n_{1}+n_{2}).
\end{equation}
Thus in the sub-Hilbert-space we write the Schr\"{o}dinger equation in the
form

\begin{equation}
H_{n,k}(t)\left| \Psi _{n,k}(t)\right\rangle _{s}=i\frac{\partial }{\partial
t}\left| \Psi _{n,k}(t)\right\rangle _{s},  \label{eq7}
\end{equation}
and $\left| \Psi (t)\right\rangle _{s}$ can be obtained from

\begin{equation}
\left| \Psi (t)\right\rangle _{s}=\sum_{n}\prod_{k}c_{n,k}\left| \Psi
_{n,k}(t)\right\rangle _{s}\left| \Phi _{k}\right\rangle  \label{eq8}
\end{equation}
where $c_{n,k}$ is time-independent and determined by the initial conditions.

For the case of this time-dependent Cini model, we set $%
A_{+}=J_{+},A_{-}=J_{-},C=J_{z}$ with $J_{\pm }=J_{1}\pm iJ_{2}$. In the
adiabatic limit, it follows from the auxiliary equations (\ref{eq23}) that

\begin{equation}
a_{i}=\theta _{i},\quad b_{i}=\varphi _{i};\quad a_{j}=\theta _{j},\quad
b_{j}=\varphi _{j}.
\end{equation}
Since

\begin{equation}
\beta _{i}=-\frac{a_{i}}{2}\exp [-ib_{i}],\quad \beta _{j}=-\frac{a_{j}}{2}%
\exp [-ib_{j}],
\end{equation}
we let both $b_{i}$ and $b_{j}$ vanish for the convenience, which leads to $%
\dot{\theta}_{i}=0$ and $\dot{\theta}_{j}=0$ in terms of the auxiliary
equations (\ref{eq23}), and then $\beta _{i}=-\frac{a_{i}}{2},\beta _{j}=-\frac{%
a_{j}}{2}.$ It is therefore easily obtained that

\begin{eqnarray}
F_{i,j}(t) &=&\exp [\frac{\lambda }{2}(\beta _{i}\beta _{j}^{\ast }-\beta
_{i}^{\ast }\beta _{j})]\left\langle j,m\right| \exp [(\beta _{j}-\beta
_{i})J_{+}-(\beta _{j}^{\ast }-\beta _{i}^{\ast })J_{-}]\left|
j,m\right\rangle  \nonumber \\
&=&\left\langle j,m\right| \exp \{[(\beta _{j}-\beta _{i})-(\beta _{j}^{\ast
}-\beta _{i}^{\ast })]J_{1}  \nonumber \\
&&+i[(\beta _{j}-\beta _{i})+(\beta _{j}^{\ast }-\beta _{i}^{\ast
})]J_{2}\}\left| j,m\right\rangle  \nonumber \\
&=&\left\langle j,m\right| \exp [i(a_{i}-a_{j})J_{2}]\left| j,m\right\rangle
,
\end{eqnarray}
where $j$ and $m$ satisfy

\begin{equation}
J_{3}\left| j,m\right\rangle =m\left| j,m\right\rangle ,\quad J^{2}\left|
j,m\right\rangle =j(j+1)\left| j,m\right\rangle .
\end{equation}
If when $t=0,$ then the state of the measuring instrument-detector
is $\left| j,j\right\rangle ,$ and the decoherence factor is
therefore
\begin{eqnarray}
F_{i,j}(t) &=&\left\langle j,j\right| \exp [i(a_{i}-a_{j})J_{2}]\left|
j,j\right\rangle  \nonumber \\
&=&[\cos (\frac{a_{i}-a_{j}}{2})]^{2j},
\end{eqnarray}
which is consistent with that obtained by Sun {\it et
al}\cite{Zeng}. For the case of classical limit where
$j\rightarrow \infty $ and $\frac{a_{i}-a_{j}}{2}\neq n\pi \quad
(n=0,\pm 1,\pm 2,\cdots ),\quad F_{k,l}(t)\rightarrow 0$, which
means the wavefunction collapse occurs under the classical limit.

Since we exactly solved the Schr\"{o}dinger equation governing the
time-dependent quantum decoherence, and the result ( i.e., the
adiabatic classical limit) obtained here is consistent with what
obtained in previous references, we hold that the more general
results presented in this paper is useful to treat the
time-dependent non-adiabatic quantum decoherence.

\section{CONCLUDING REMARKS}

The present paper obtains exact solutions and decoherence factor
of the general time-dependent dynamical model for quantum
decoherence by working in the sub-Hilbert space corresponding to
the eigenvalue of two invariants and by making use of the
invariant-related unitary transformation method. The
invariant-related unitary transformation formulation is an
effective method for treating time-dependent
problems\cite{Fu,Shen2,Shen3}. This formulation replaces
eigenstates of the time-dependent invariants with those of the
time-independent invariants through the unitary transformation. It
uses the invariant-related unitary transformation and obtains the
explicit expression for the time-evolution operator, instead of
the formal solution that is related to the chronological product.
In view of what has been discussed above, it can be seen that the
invariant theory is appropriate to treat the time-dependent
quantum decoherence. Apparently, the results presented in the
present paper is easy to generalize to the time-dependent
Hepp-Coleman model. Since the geometric phase factor appears in
time-dependent systems, it is interesting to investigate the
geometric phase in the time-dependent quantum decoherence by using
the formulation in this paper.

Acknowledgment%
%
This project was supported by the National Natural Science
Foundation of China under the project No. $30000034$.

\end{document}